\documentclass[journal=jacsat,manuscript=article]{achemso}
\usepackage{chemformula} % Formula subscripts using \ch{}
\usepackage[T1]{fontenc} % Use modern font encodings
\usepackage{amsfonts}    % Blackboard math
\usepackage{bm}
\usepackage{subfig}
\usepackage{subcaption}
\usepackage{pdflscape}
\usepackage{hyperref}
\usepackage{lineno}
\usepackage{float}
\usepackage{longtable}
\usepackage{xr}

\SectionNumbersOn
\hypersetup{
  colorlinks   = true, %Colours links instead of ugly boxes
  urlcolor     = blue, %Colour for external hyperlinks
  linkcolor    = black, %Colour of internal links
  citecolor    = black, %Colour of citations
  breaklinks   = true
}
\author{AnnaElaine L. Rosengart}
\email{arosenga@andrew.cmu.edu}
\affiliation[Carnegie Mellon University]
{Department of Statistics and Data Science, Carnegie Mellon University}

\author{Amanda L. Bidwell}
\affiliation[Stanford University]
{Department of Civil and Environmental Engineering, School of Engineering and Doerr School of Sustainability, Stanford University}

\author{Marlene K. Wolfe}
\affiliation[Emory University]
{Gangarosa Department of Environmental Health, Rollins School of Public Health, Emory University}

\author{\newline Alexandria B. Boehm}
\affiliation[Stanford University]
{Department of Civil and Environmental Engineering, School of Engineering and Doerr School of Sustainability, Stanford University}

\author{F. William Townes}
\affiliation[Carnegie Mellon University]
{Department of Statistics and Data Science, Carnegie Mellon University}

\title[]
  {Spatio-Temporal Variability of the Pepper Mild Mottle Virus Biomarker in Wastewater}

\abbreviations{PMMoV, WBE, WWTP}

\begin{document}

% \begin{tocentry}

%   \comment{TO DO: make graphical abstract}

%   Some journals require a graphical entry for the Table of Contents.
%   This should be laid out ``print ready'' so that the sizing of the
%   text is correct.

%   Inside the \texttt{tocentry} environment, the font used is Helvetica
%   8\,pt, as required by \emph{Journal of the American Chemical
%   Society}.

%   The surrounding frame is 9\,cm by 3.5\,cm, which is the maximum
%   permitted for  \emph{Journal of the American Chemical Society}
%   graphical table of content entries. The box will not resize if the
%   content is too big: instead it will overflow the edge of the box.

%   This box and the associated title will always be printed on a
%   separate page at the end of the document.

% \end{tocentry}

\begin{abstract}
  Since the start of the coronavirus-19 pandemic, the use of wastewater-based epidemiology (WBE) for disease surveillance has increased throughout the world. Because wastewater measurements are affected by external factors, processing WBE data typically includes a normalization step in order to adjust wastewater measurements (e.g.\ viral RNA concentrations) to account for variation due to dynamic population changes, sewer travel effects, or laboratory methods. Pepper mild mottle virus (PMMoV), a plant RNA virus abundant in human feces and wastewater, has been used as a fecal contamination indicator and has been used to normalize wastewater measurements extensively. However, there has been little work to characterize the spatio-temporal variability of PMMoV in wastewater, which may influence the effectiveness of PMMoV for adjusting or normalizing WBE measurements. Here, we investigate its variability across space and time using data collected over a two-year period from sewage treatment plants across the United States. We find that most variation in PMMoV measurements can be attributed to longitude and latitude followed by site-specific variables. Further research into cross-geographical and -temporal comparability of PMMoV-normalized pathogen concentrations would strengthen the utility of PMMoV in WBE.
\end{abstract}

%% SYNOPSIS %%%%%%%%%%%%%%%%%%%%%%%%%%%%%%%%%%%%%%%%%%%%%%%%%%%%%%%%%%%%%%%%
\section{Synopsis}
\label{sec:synopsis}
PMMoV is a widely used proxy for human fecal content used to normalize pathogen measurements in WBE. We investigate its spatio-temporal variability using nationwide data from mid-2021 through mid-2023 from United States sewage treatment plants. 

%% INTRODUCTION %%%%%%%%%%%%%%%%%%%%%%%%%%%%%%%%%%%%%%%%%%%%%%%%%%%%%%%%%%%%%%%%
\section{Introduction}
\label{sec:introduction}
Wastewater-based epidemiology (WBE) is now a well established method of surveilling community health over a large area \cite{kirby2021, isaksson2022}. For certain viral illnesses, such as COVID-19 and Influenza, infected individuals shed pathogenic genetic material (analyte) in their feces, urine, and sputum, which then enters a sewer system \cite{zhan2022}. Quantitative and droplet digital (reverse transcription-) polymerase chain reaction (PCR) are cost- and time-effective methods for absolute quantification of viral genetic material in a wastewater sample \cite{diamond2022}. Under the assumption that changes in the number of viral gene copies determined by PCR are reflective of changes in the number of infections in an area, these measurements from wastewater can be used to infer trends in community disease burden \cite{li2021,weidhaas2021,wolfe2021,duvallet2022}.

However, wastewater measurements are subject to other sources of variation: in-human, in-sewer, and in-lab effects \cite{boyce2023}. In-human effects occur prior to an analyte's entry into the sewer system and may be caused by changes in population (e.g.\ for a sporting event) or fecal shedding rates for different strains of a virus. In-sewer effects occur during sewer travel and include dilution due to groundwater infiltration, degradation of genetic material caused by fluctuating temperature or pH, or the adsorption rate of genetic material to solid waste. In-lab effects occur during sampling and analysis and can be caused by differences in sampling (e.g.\ solid vs.\ liquid samples) and laboratory protocols (e.g.\ freezing and transport of samples, instrument bias, nucleic acid extraction efficiency). These sources of variability are due to the environment, not necessarily disease dynamics, and may introduce noise that reduces the correlation between wastewater measurements and disease incidence (or prevalence, though evidence suggests shedding, at least in the case of SARS-CoV-2, is greatest in the early stages of infection, making incidence more relevant \cite{arts2023}) \cite{boyce2023}. 

Normalizing concentrations attempts to account for these effects to improve comparisons across space and time. Metadata, such as the catchment population size or the flow rate of water entering a wastewater treatment facility (site), can be used to adjust for in-human effects and certain in-sewer effects like groundwater infiltration. For example, flow normalized concentration can be calculated as:
\begin{equation*}
\frac{\frac{\text{analyte gene copies}}{\text{1 liter}} \cdot \frac{\text{flow in liters}}{\text{1 day}}}{\text{sewershed population}}= \text{analyte gene copies} \text{ per person per day}
\end{equation*}
Days with greater precipitation may yield smaller raw concentrations due to dilution, not a decrease in cases. Multiplying by flow rate can adjust for this sewer effect and thereby improve the ability of the wastewater measurements to represent disease incidence. However, metadata normalization cannot adjust for in-lab effects because the information it uses is not intrinsic to a sample. 

Biomarkers used for normalization are biological agents present in a sample and are thought to enter the sewer system and be affected by wastewater and laboratory methods in ways similar to the analyte of interest. Biomarker normalization is performed by calculating the ratio of an analyte's concentration to that of the chosen biomarker, for example:
\begin{equation*}
\frac{\text{analyte gene copies / gram}}{\text{marker gene copies / gram}} = \text{normalized analyte value}
\end{equation*}
Though the result is a unitless measure, normalization by biomarkers has the potential to correct for all three types of environmental variation \cite{boyce2023, holm2022, hsu2022, maal2023}.

A popular biomarker for wastewater normalization is pepper mild mottle virus (PMMoV), a virus infecting the genus \emph{Capsicum}, including bell and spicy peppers, that appears in human fecal matter after the consumption of infected plants. PMMoV has previously been used as an indicator for the presence or absence of fecal contamination in ocean water, river water, and treated wastewater \cite{rosario2009, hamza2011, symonds2016, bivins2020} and exhibits properties that give it promise as a normalizing agent. It is one of the most abundant RNA viruses found in human feces, making it easily quantifiable \cite{zhang2005}. Though it is non-enveloped, PMMoV is a single-stranded RNA virus like SARS-CoV-2 and many other viruses \cite{kitajima2018, mariano2020}, and therefore it may experience effects through sewer system travel similar to that of these pathogens of interest \cite{kitajima2018, bivins2020, holm2022}. However, its utility in practice has been relatively inconclusive. Several studies have shown that pathogen concentrations normalized with PMMoV have improved correlations with disease incidence compared to raw concentrations \cite{ai2021, zhan2022}. \citet{wolfe2021} showed through a mass-balance model that concentrations of SARS-CoV-2 RNA scaled by PMMoV RNA were directly proportional to COVID-19 incidence rate. In contrast, others have reported only mild increases or decreases in correlations \cite{feng2021, greenwald2021, duvallet2022, maal2023}. 

PMMoV concentrations may differ throughout the year and across different regions, and this may confound its use as a WBE normalizer and explain some of the mixed results in the literature. PMMoV concentrations have been found to be relatively stable in comparison to other human fecal indicators \cite{greenwald2021} but also to vary greatly both across regions and over time \cite{haramoto2013, symonds2016, dhakar2022}. One study reported PMMoV concentrations to exhibit little evidence of seasonal trend \cite{holm2022} while another showed mild seasonality \cite{dhiyebi2023}. Even within a single individual, PMMoV shedding rates can vary greatly over time \cite{arts2023}. 

It is this uncertainty in the characteristics of PMMoV concentration over space and time that motivate this work. We use data from the ongoing wastewater sampling project, WastewaterSCAN \cite{wastewaterscan2023}, and fit four models that incorporate geographical, temporal, and site-specific information in order to investigate three questions of interest: (i) how does PMMoV concentration vary across locations; (ii) how does it vary over time; and (iii) what proportion of its variation can be accounted for solely by these spatio-temporal factors. We are able to quantify the variance explainable, visualize the trends in PMMoV concentration based upon variables of interest, and suggest potential contributors to the variation.

%% METHODS %%%%%%%%%%%%%%%%%%%%%%%%%%%%%%%%%%%%%%%%%%%%%%%%%%%%%%%%%%%%%%%%%%%%%
\section{Materials and Methods}
\label{sec:methods}

\subsection{Data}
\label{subsec:data}
The data comprise PMMoV concentrations taken from wastewater samples across the United States and collected as part of the WastewaterSCAN project \cite{wastewaterscan2023}. Viral RNA copies were quantified from wastewater solids using reverse transcription droplet digital PCR. These extraction and quantification procedures have previously been described in detail \cite{topol2021a, topol2021b, topol2022}. Additionally, the full methods can be found in two data descriptors \cite{boehm2023, boehm2024}. 

\begin{table}[!htb]
  \centering
  \small
  \caption{Summary statistics for the number of observations per site, the PMMoV concentration across all sites, and the average daily precipitation across all sites (gc = gene copies; g = grams; in = inches).}
  \begin{tabular}{l|c|c|c|}
       & \multicolumn{1}{|p{2.8cm}|}{\centering Number of \\ Observations} 
       & \multicolumn{1}{|p{2.8cm}|}{\centering PMMoV \\ ($\log_{10}$ gc/g)} 
       & \multicolumn{1}{|p{3.0cm}|}{\centering Average \\ Precipitation (in)} 
       \\ \hline
       Min. & 30  & 5.92             & 0.00            \\
       Max. & 812 \hspace{1.2mm} & 11.31 \hspace{1.4mm} & 4.34  \\
       Med. & 112 \hspace{1.2mm} & 8.78             & 0.00   \\
       Mean. & \hspace{0.9mm} 158.64 & 8.77  & 0.10
  \end{tabular}
  \label{tab:variable-summary}
\end{table}
Sites with at least 30 samples collected from May 29th, 2021, through August 18th, 2023, were included in the analysis for a total of 25,383 observations from 160 sites across 31 states. Each site was classified as having a separated or combined sewer system depending upon whether the system accepted sanitary and runoff water together. Precipitation data were collected from the Global Historical Climatology Network daily database from the National Centers for Environmental Information of the National Oceanic and Atmospheric Administration \cite{noaa2023}. The average precipitation by day was calculated for each site by taking the mean of daily measurements over all stations located in all counties served by the plant. Latitude and longitude data were taken as the centroid of the zipcode associated with each site. Most sites had over 100 samples, and most samples were taken on days with 0 inches of precipitation (Table~\ref{tab:variable-summary}; see Supplemental Table~\ref*{tab:summary-sites} for summary statistics by site).

\subsection{Modeling}
\label{subsection:modeling}
We fit four different models that describe the conditional distribution of $\log_{10}$ PMMoV concentration using a linear combination of spatio-temporal factors. We use the following notation:  $\text{lat}_i$ = latitude of site $i$ in degrees; $\text{lng}_i$ = longitude of site $i$ in degrees; $\text{sewer}_i$ = sewer system type of site $i$ (1 for combined and 0 for separated); $\text{prcp}_{i, t}$ = average precipitation $t$ days after May 28th, 2021, at site $i$ (in inches); $\text{log}_{10} \text{PMMoV}_{i, t}$ is the $\log_{10}$ PMMoV concentration $t$ days after May 28th, 2021, at site $i$ (in gene copies per gram dry weight); $\text{site ID}_i$ is an indicator value for site $i$. 

\subsubsection{Simple Median Model}
\label{subsubsection:model-1}
The simple median model is a quantile regression model for the median $\log_{10}$ PMMoV concentration:
\begin{equation}
  \label{eq:model-1}
  Q_{0.5}(\text{log}_{10} \text{PMMoV}_{i}) = \beta_0 + \beta_1 \cdot \text{lat}_i + \beta_2 \cdot \text{lng}_i
\end{equation} 
where $Q_\tau(\cdot)$ denotes the $\tau$-th quantile of the distribution of the random variable of interest conditional on some set of covariates, which are taken here to be latitude and longitude. This model was used for investigating the variation in PMMoV concentration solely on the basis of the geographic origin of a sample. 

\subsubsection{Detailed Median Model}
\label{subsubsection:model-2}
The detailed median model expands upon the simple median model by including average daily precipitation, sewer system type, and their interaction. Fourier basis functions with week- and year-long periods are also included, motivated by the potential seasonality in pepper consumption as well as evidence of a weekly pattern in the autocorrelation of the data in exploratory analysis (Supplemental Figure~\ref*{fig:autocorr-plot-data}). 

Each pair of basis functions is defined as:
\begin{equation}
  \label{eq:basis-funcs}
  \psi^{\text{sin}}_{\lambda}(t) = \frac{\sin(\frac{2\pi t}{\lambda})}{\sqrt{\lambda / 2}} 
  \hspace{15mm}
  \psi^{\text{cos}}_{\lambda}(t) = \frac{\cos(\frac{2\pi t}{\lambda})}{\sqrt{\lambda / 2}}
\end{equation}
where $\lambda$ corresponds to the period of the bases in days. For weekly trends we set $\lambda = 7$, and for annual trends we set $\lambda = 365.25$. This model assumes the form:
\begin{equation}
  \label{eq:model-2}
  \begin{split}
    Q_{0.5}(\text{log}_{10} \text{PMMoV}_{i, t}) &= \beta_{0} + \beta_1 \cdot \text{lat}_i + \beta_2 \cdot \text{lng}_i \\
    &+ \beta_3 \cdot \text{prcp}_{i,t} + \beta_4 \cdot \text{sewer}_i + \beta_5 \cdot \text{prcp}_{i,t} \cdot \text{sewer}_i \\
    &+ \beta_6 \cdot \psi^{\text{sin}}_{7}(t) + \beta_7 \cdot \psi^{\text{cos}}_{7}(t) \\
    &+ \beta_{8} \cdot \psi^{\text{sin}}_{365.25}(t) + \beta_{9} \cdot \psi^{\text{cos}}_{365.25}(t)
  \end{split}
\end{equation}
To guard against model misspecification, we used the cluster-robust bootstrap for inference, which does not require assumptions on the distribution of the errors \cite{hagemann2017}.

\subsubsection{Variance Decomposition Model}
\label{subsubsection:model-3}
We fit a variance decomposition model in order to attribute portions of the variability to spatial, temporal, and site-specific sources by partitioning the model's coefficient of determination, $R^2$. This model assumes the conditional mean, rather than the conditional median, of $\log_{10}$ PMMoV can be described as a linear combination of the chosen covariates, and that the errors have finite variance (but not necessarily that they are normally distributed):
\begin{equation}
  \label{eq:model-3}
  \begin{split}
    \mathbb{E}\left[ \text{log}_{10} \text{PMMoV}_{i, t} \right] &= \beta_{0} + \beta_1 \cdot \text{lat}_i + \beta_2 \cdot \text{lng}_i + \beta_3 \cdot \text{prcp}_{i,t} + \beta_4 \cdot \text{sewer}_i \\
    &+ \beta_5 \cdot \text{prcp}_{i,t} \cdot \text{sewer}_i + \beta_6 \cdot \text{site  ID}_i \\
    &+ \beta_7 \cdot \psi^{\text{sin}}_{7}(t) + \beta_8 \cdot \psi^{\text{cos}}_{7}(t) + \beta_{9} \cdot \psi^{\text{sin}}_{365.25}(t) + \beta_{10} \cdot \psi^{\text{cos}}_{365.25}(t)
  \end{split}
\end{equation}
We obtained the partition by successively adding groups of covariates until reaching the final model described in Equation~\ref{eq:model-3}, recording the $R^2$ at each step. These values represent the percent variation explained by the newly added covariates that remained unexplained by the covariates in the previous, smaller model fit. The order of covariate addition was: (i) latitude and longitude; (ii) precipitation, sewer system type, and their interaction; (iii) site indicator; (iv) the weekly and yearly time components.

\subsubsection{Bayesian Median Model}
\label{subsubsection:model-4}
For each site separately, we fit a Bayesian model using precipitation and both weekly and yearly time components. Formally, we assume for each site independently:
\begin{equation}
  \label{eq:model-4}
  \begin{split}
    \text{log}_{10} \text{PMMoV}_{t} &= 
      \beta_{0} + \beta_1 \cdot \text{prcp}_{t} + \beta_2 \cdot \psi^{\text{sin}}_{7}(t) + \beta_3 \cdot \psi^{\text{cos}}_{7}(t) \\
      &+ \beta_4 \cdot \psi^{\text{sin}}_{365.25}(t) + \beta_5 \cdot \psi^{\text{cos}}_{365.25}(t) + \epsilon_{t}
  \end{split}
\end{equation}
where $\epsilon_{t}$ are independent and identically distributed Laplace random variables with location of 0 and scale of $\sigma$. Maximum likelihood estimation under the assumed Laplace distribution is equivalent to minimization of the quantile regression loss function, making these conditional median models \cite{geraci2006}. 

Inference was done with Hamiltonian Monte Carlo using the No-U-Turn Sampler variant \cite{hoffman14, betancourt2018}. The intercept, $\beta_0$, was given a uniform prior over the real line, while every other $\beta_i$ was given a Gaussian prior with 0 mean and standard deviation $\sigma_i$. We used a Student's $t$ prior with 5 degrees of freedom, location of 0, and scale of 1 for $\sigma$ as well as each $\sigma_i$. Further details on the fitting regime can be found in the supplementary code scripts.

We used these models to visually describe the temporal variation in PMMoV concentration by comparing the model predictions with the observed values. Fitting to each site separately afforded us the ability to investigate differences in the effects of our chosen covariates for different sites.

\subsection{Data Analysis}
\label{subsection:data-analysis}
Data analyses were performed in R version 4.1.0 (Camp Potenzen) \cite{r2021} using RStudio version 2024.04.0+735 \cite{rstudio2020}. The simple and detailed median models were fit using the \emph{quantreg} package \cite{koenker2022}, and the cluster-robust wild bootstrap with site membership grouping was used for uncertainty quantification \cite{hagemann2017}. The variance decomposition model was fit with the \emph{stats} package \cite{r2021}, and the Bayesian median models were fit with the \emph{rstan} package \cite{rstan2024}. A detailed list of packages used in the analyses can be found in the supplementary information. 

%% RESULTS %%%%%%%%%%%%%%%%%%%%%%%%%%%%%%%%%%%%%%%%%%%%%%%%%%%%%%%%%%%%%%%%%%%%%
\section{Results and Discussion}
\label{sec:results}

\subsection{Geographic Variation in PMMoV Concentration}
\label{subsec:geo-var}
\begin{landscape}
  \begin{figure}
   \centering
   \includegraphics[height=0.7\textwidth]{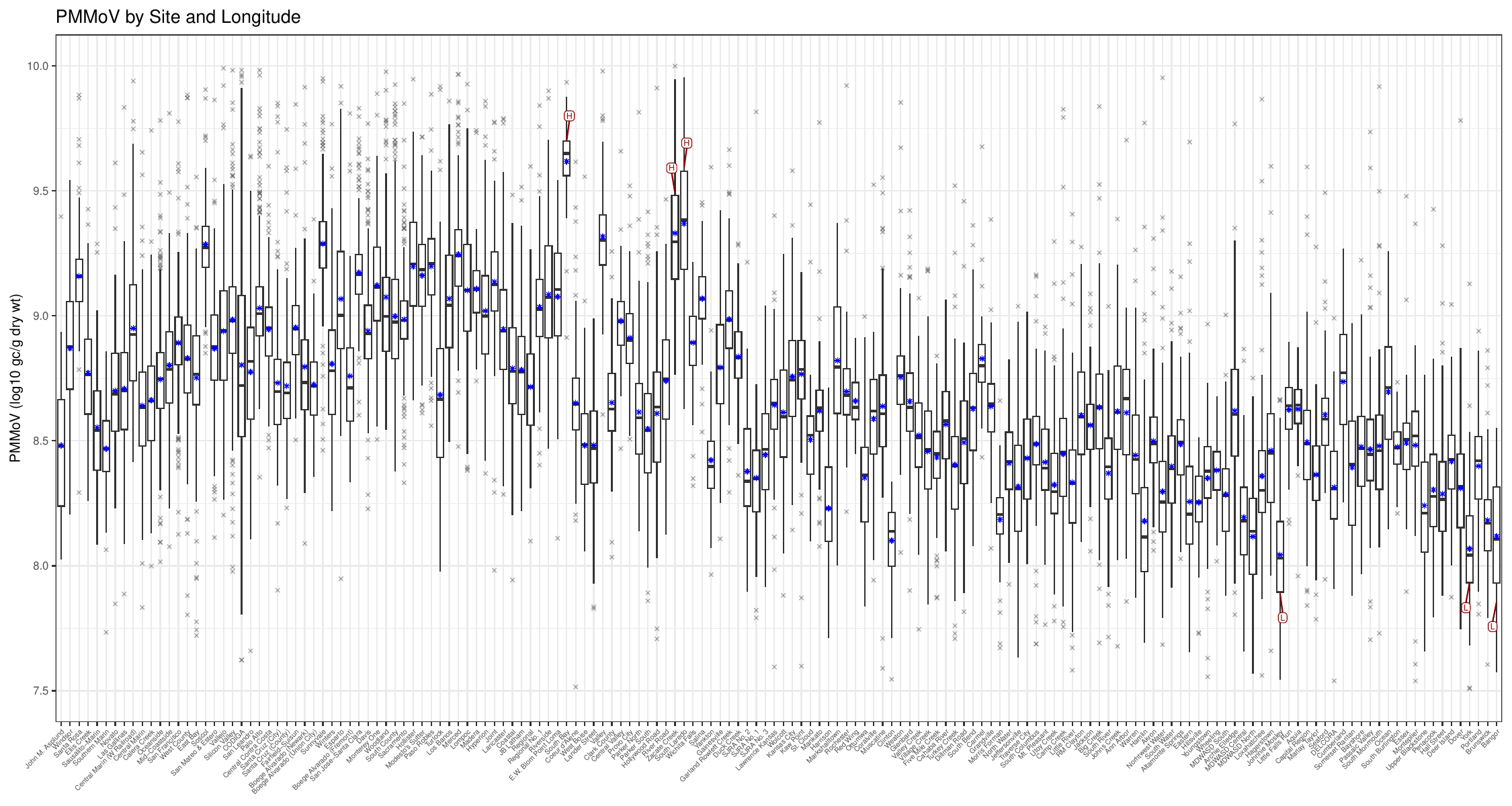}
   \caption{PMMoV concentration varies within and between sites. Sites ordered by longitude (left to right = west to east). Top and bottom of each box demarcate the interquartile range (75th and 25th percentiles, respectively) for concentrations taken from each site. Black horizontal lines and blue stars mark site median and mean concentrations, respectively. Observations falling outside the vertical whiskers have values exceeding 1.5 times the interquartile range and are marked with a grey \emph{x}. The three sites with the highest median concentrations are marked with a red \emph{H}, and the three sites with the lowest medians are marked with a red \emph{L}. Vertical axis limited to 7.5 to 10.0 gc/g dry wt for visibility. See Supplemental Table~\ref*{tab:abbrev} for site name abbreviations. (gc = gene copies; g = gram; wt = weight)}
   \label{fig:box-whisker}
  \end{figure}
\end{landscape}

PMMoV concentration varies widely both within and across sites. The majority of sites have concentrations spanning at least one order of magnitude, and even the smallest interquartile range (South Burlington-Airport Parkway WWTF) ranges from 8.41 to 8.54 $\log_{10}$ gene copies per gram dry weight. Median concentration tends to decrease with increasing site longitude (i.e.\ moving east), and sites located in the west experience concentrations that are generally higher, on average, than those experienced by sites in the midwest and east (Figure~\ref{fig:box-whisker}). 

\begin{figure}
  \centering
  \includegraphics[width=\textwidth]{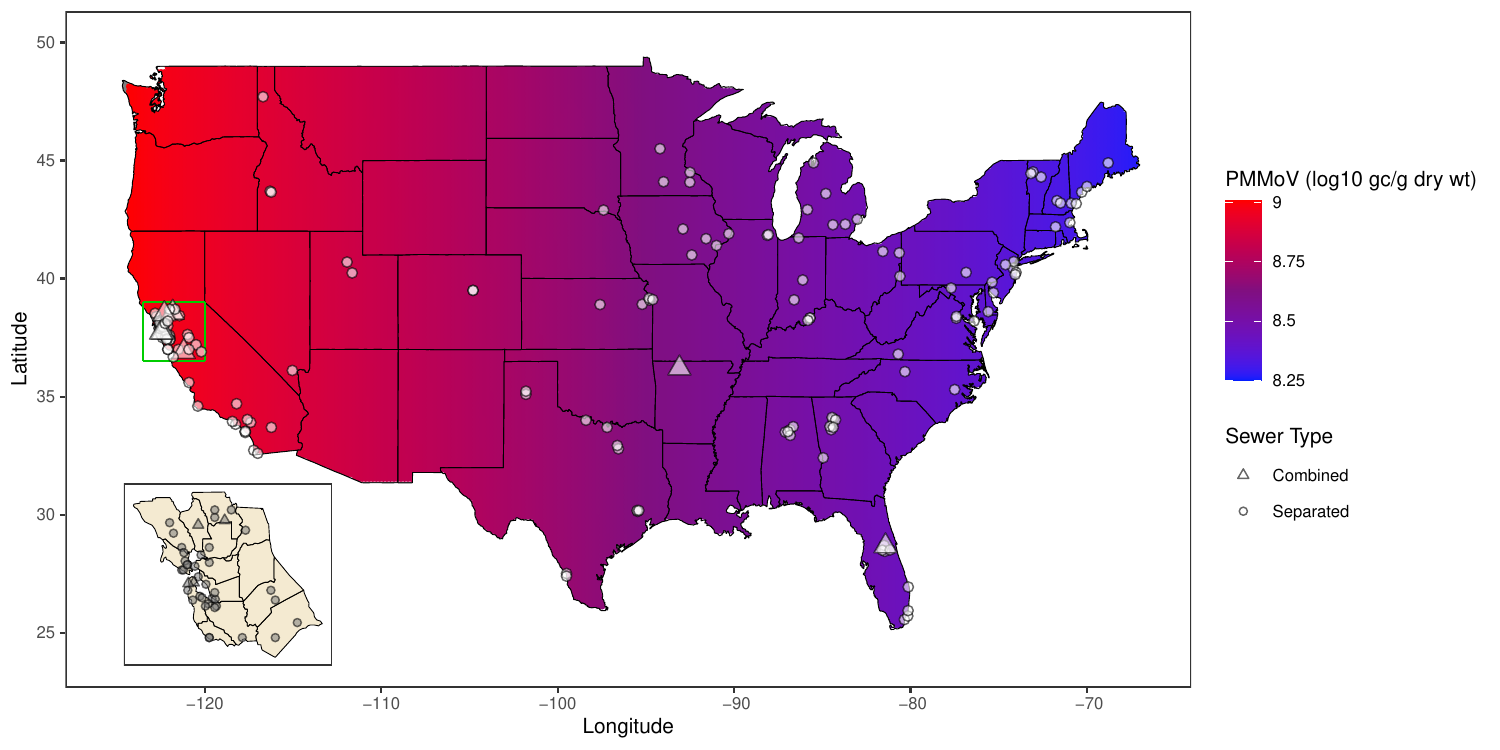}
  \caption{Median PMMoV concentration tends to be greater in western regions compared to northeastern areas in the contiguous United States. Sampled site locations as points shaped by sewer type are superimposed on a color gradient illustrating the concentration predicted by the simple median model (main). Sites in the Bay Area and nearby counties in higher resolution (inset) on a neutral background. Alaska omitted for visibility. (gc = gene copies; g = gram; wt = weight)}
  \label{fig:us-site-map}
\end{figure}

The color shift when moving across the map reiterates the strong association between longitude and PMMoV concentration (Figure~\ref{fig:us-site-map}). This result is confirmed by the statistical significance in the coefficient on longitude ($\beta = -1.29 \times 10^{-2}$; $p < 1.00 \times 10^{-15}$) of the simple median model. The negative coefficients on both latitude and longitude (Supplemental Table~\ref*{tab:model-1}) indicate that predicted $\log_{10}$ PMMoV concentration decreases with more northern and more eastern sampling locations. Even when including additional covariates, as in the detailed median model, the sign and significance of longitude remain ($\beta = -1.32 \times 10^{-2}$; $p < 1.00 \times 10^{-15}$), and the coefficient for latitude also remains negative but is not significant.

The three sites with the lowest median concentrations are Johnnie Mosley Regional Water Reclamation Facility in Kinston, NC; York Sewer District in York Beach, ME; and the City of Bangor Wastewater Treatment Plant in Bangor, ME (Figure~\ref{fig:box-whisker}, \emph{L}s). All three of these sites are located on the eastern coast of the United States. The three sites with the highest median concentrations are South Bay International Wastewater Treatment Plant in San Diego, CA; South Laredo Wastewater Treatment Plant in Laredo, TX; and Zacate Creek Wastewater Treatment Plant in Laredo, TX (Figure~\ref{fig:box-whisker}, \emph{H}s). South Laredo and Zacate Creek are located along the Rio Grande River on the Mexico border, and South Bay treats sewage from Tijuana, Mexico \cite{cali2024}. We hypothesize that this variation may be driven by geographical differences in diet. Pepper mild mottle virus appears in human feces through the ingestion of infected plant products such as salsa, hot sauce, and spices \cite{rosario2009, zhang2005}. Regional popularity of these foods, or even agricultural practices that affect the availability of pepper products, may therefore affect its presence in wastewater.

\subsection{Variance Partition}
\label{subsec:var-part}
\begin{figure}[H]
  \centering
  \includegraphics[width=\textwidth]{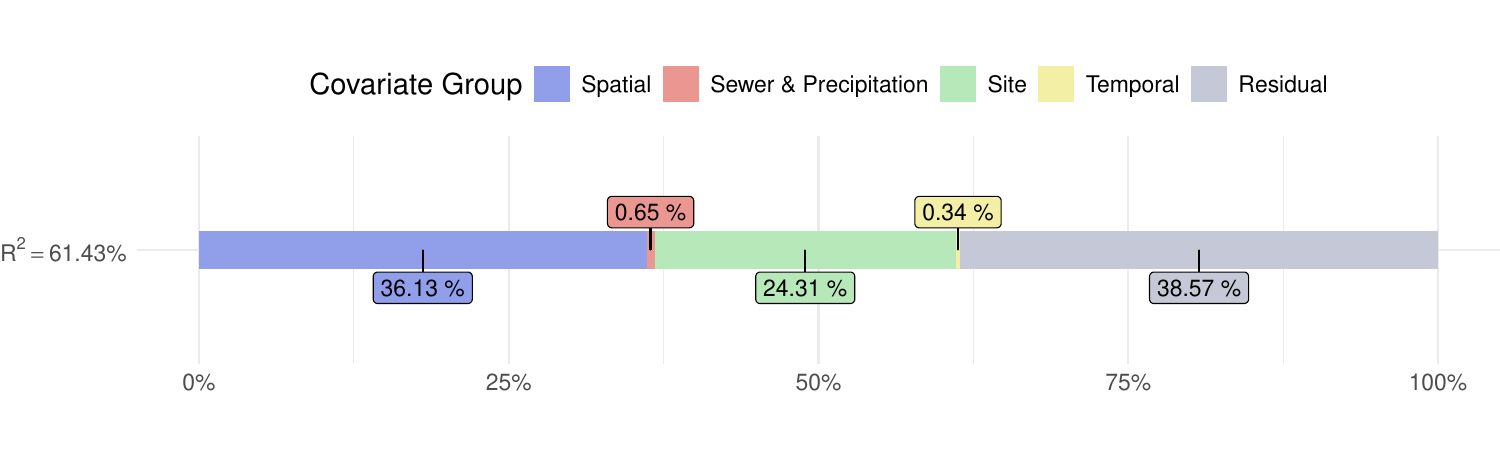}
  \caption{Location and site membership account for the majority of the variation in PMMoV concentration. The spatial components (latitude and longitude) account for the greatest portion at about 36\%. Site membership accounts for over 24\% of the variation that remains unexplained by the spatial, sewer, and precipitation variables. The temporal components, sewer system type, and precipitation altogether account for about 1\% of the variation.}
  \label{fig:var-decomp}
\end{figure}

The majority of the variation in PMMoV concentration can be accounted for by spatial variables and an indicator of site membership (Figure~\ref{fig:var-decomp}). Moreover, geographic location (latitude and longitude) of a sample as well as site membership are the two greatest sources of variation in PMMoV concentration. These two groups of covariates explain over 36\% and 24\% of the variance, respectively, in the variance decomposition model, which agrees with the visual and statistical results from the simple median and detailed median models. 
Precipitation, sewer type, and temporal variables account for some of the variability as well, although the portion is quite small in comparison to geography and site membership. The remaining variance (Figure~\ref{fig:var-decomp}, grey bar) is within-site variation that is unaccounted for by our chosen covariates. This may be due to microscale variation or noise; we are unable to investigate further due to data limitations. The interaction of the temporal components with the spatial and site indicator terms was also investigated, but these features accounted for only around 2\% of the variance (Supplemental Figure~\ref*{fig:var-decomp-interactions}).

Recent work has found evidence that factors, such as alkalinity, biochemical oxygen demand, and flow rate, can affect PMMoV concentrations through dilution, degradation, and adsorption of genetic material to solids during sewer travel. Even for sites serving the same city, physico-chemical parameters can have statistically significant differences in levels \cite{goitom2024}. Thus, the large portion of the variation attributable to site membership in Figure~\ref{fig:var-decomp} may be due to the physical and chemical attributes of the sewer system sampled.

\subsection{Temporal Variation}
\label{subsec:temp-var}
\begin{figure}[H]
  \centering
  \includegraphics[width=\textwidth]{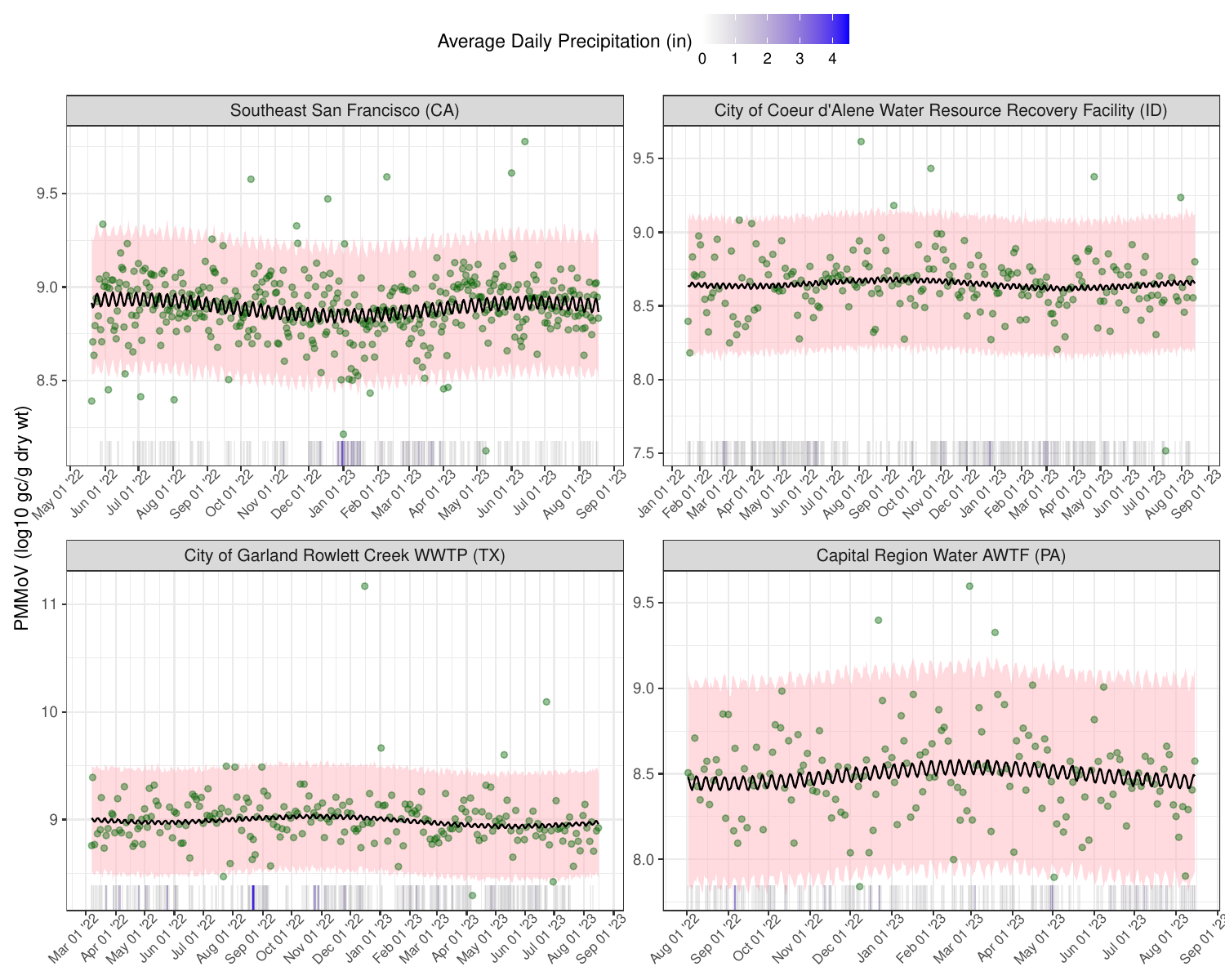}
  \caption{PMMoV concentration varies over time, both on a weekly and yearly scale, and in different ways for different sites. Black line is the median predicted PMMoV of 4,000 posterior samples. Lower and upper limits of pink ribbon are 2.5\% and 97.5\% quantiles, respectively, from the samples. Green points show observed concentrations for available dates. Rug shows average daily precipitation. Sites were chosen to illustrate results for different US regions. (gc = gene copies; g = gram; wt = weight; in = inches)}
  \label{fig:predictions}
\end{figure}
Patterns in day-to-day and seasonal changes in PMMoV concentration vary by site. Weekly variation is reflected in the tight oscillation of the median predicted concentrations (Figure~\ref{fig:predictions}); however, the proportion of sites with statistically significant weekly terms was quite small (12/160 for the coefficient on $\psi^{\text{sin}}_{7}$ and 23/160 for the coefficient on $\psi^{\text{cos}}_{7}$; Supplemental Tables~\ref*{tab:sinwk} and~\ref*{tab:coswk}).

Seasonality is illustrated in the wider sinusoidal pattern (Figure~\ref{fig:predictions}). For example, Southeast San Francisco in California sees generally higher concentrations in the summer, while greater concentrations occur in late winter for Capital Region Water AWTF in Pennsylvania. Moreover, a larger proportion of yearly terms were statistically significant (51/160 for the coefficient on $\psi^{\text{sin}}_{365.25}$ and 42/160 for the coefficient on $\psi^{\text{cos}}_{365.25}$; Supplemental Tables~\ref*{tab:sinyr} and~\ref*{tab:cosyr}). 

This temporal variability is in agreement with some prior research into the variation of PMMoV over time \cite{chettleburgh2023} but contrasts with others that did not report evidence of seasonal patterns \cite{kitajima2018, hamza2011}.Additional long-term data collection of PMMoV concentrations would be beneficial for defining this time-based variation with more certainty.

\subsection{Effect of Precipitation}
\label{subsec:precipitation}
\begin{figure}
  \centering
  \includegraphics[width=0.9\textwidth]{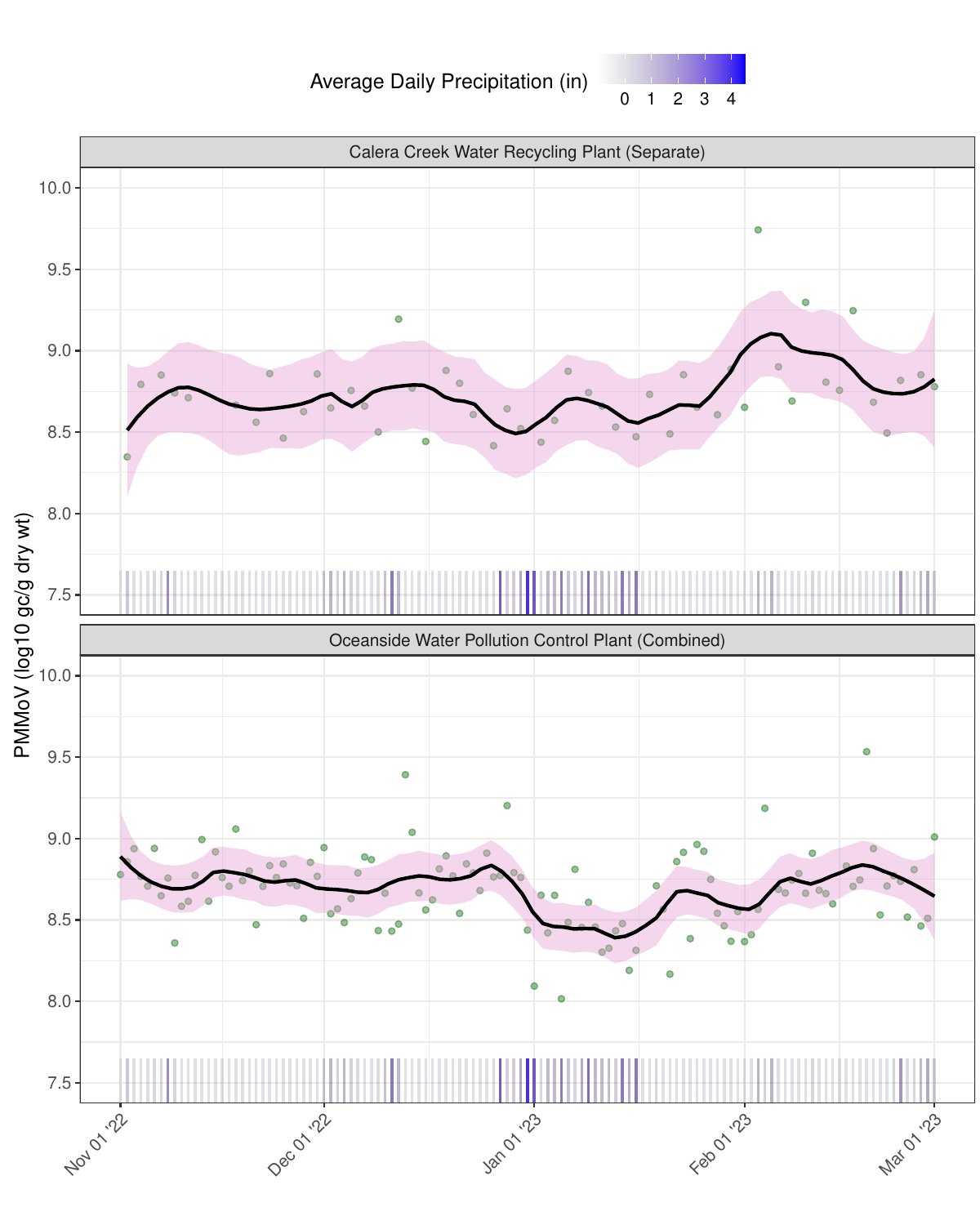}
  \caption{PMMoV concentration remains constant at Calera Creek (separated system) and decreases at Oceanside (combined system) over a period of increased precipitation. Black line is loess smoother with 95\% confidence interval. Green points show observed concentrations for available dates. Rug shows average daily precipitation. (gc = gene copies; g = gram; wt = weight; in = inches)}
  \label{fig:sewer}
\end{figure}

PMMoV concentration drops on days of high precipitation for sites with combined sewer systems. Calera Creek Water Recycling Plant and Oceanside Water Pollution Control Plant are both located in the Bay Area of central California and serve San Mateo County. Due to their proximity, the sewer catchments of both sites experience similar levels of precipitation. However, during a period of higher precipitation, PMMoV concentration at Oceanside, which has a combined sewer system, decreased while it remained relatively constant at Calera Creek, which has a separated system (Figure~\ref{fig:sewer}). 

When considering all sites together as in the detailed median model, the coefficient for precipitation is statistically significant ($\beta = -7.33 \times 10^{-2}$; $p = 1.22 \times 10^{-13}$). The coefficients on the sewer type term and the interaction between precipitation and sewer type are also negative though not significant (Supplemental Table~\ref*{tab:model-2}), potentially due to the small sample size of combined sewer systems in our dataset. The coefficient signs imply that greater levels of precipitation are associated with lower concentrations of PMMoV; sites with combined sewer systems experience lower concentrations compared to those with separated systems; and a combined sewer system modifies the effect of precipitation such that PMMoV decreases more. 

Greater amounts of precipitation can lead to dilution effects from groundwater infiltration \cite{li2021}, thus leading to lower measured concentrations of viral nucleic acids. In addition, sewer system type mediates the entrance of precipitation into a wastewater treatment plant by allowing dilution by runoff. \citet{goitom2024} reported a negative association between PMMoV concentration and precipitation for one site included in their study, which they proposed was due to the combined sewers in the served city. However, it has also been suggested that periods of high precipitation can lead to higher viral concentrations, potentially due to the scouring of solid material in the sewers by the higher flow rates or due to the expedited travel time which would reduce degradation of genetic material \cite{goitom2024,bertels2022}. This may provide an explanation for why not all sites have negative coefficients for precipitation in the individual Bayesian model fits (Supplemental Table~\ref*{tab:prcp}).

\section{Conclusion}
We describe the majority of the variation in PMMoV concentration across a large sample of United States sewer treatment plants using spatio-temporal factors. Our work provides insights into the temporal and geographical trends in PMMoV concentration, showing that it has high levels of variation both within and across sites. Differences across sites may be due to features of the wastewater matrix or sewer system at different facilities, while differences within sites may be due to weather changes or variation in pepper consumption.  

Some site-to-site differences and the changes associated with fluctuating precipitation levels suggest PMMoV normalization is effective at accounting for many in-human and in-sewer effects. However, the longitudinal variation may negatively impact its performance as a normalizer. For example, a location in the southwest US and a location in the northeast US experiencing similar burdens of disease could see large differences in normalized concentrations due to the fact that southwestern sites have, on average, higher concentrations of PMMoV. We hypothesize that this trend may be due to patterns in diet, and future work should consider incorporating information about pepper product consumption in order to further improve correlations. 

Our work has several limitations that may affect our findings and their generalizability. These include having very few sites with combined sewer systems, an overrepresentation of data from California sites, and no data from outside the COVID-19 pandemic. Quantification was from only solid samples, which enabled us to better investigate in-human and in-sewer effects by reducing confounding from in-lab effects. Consequently, we did not investigate in-lab effects, but others have found PMMoV to be useful in adjusting for between-sample variation from laboratory processing \cite{wu2020}. Future studies should examine whether these results hold for different sampling methods (grab vs.\ composite, time- vs.\ flow-proportional, liquid vs.\ solid), which have been shown to affect viral concentrations \cite{li2021}. Replication of this study with longer time series and more participating treatment facilities would provide additional insights. However, datasets of this size containing raw concentrations rather than smoothed, pre-normalized values or summary statistics are not publicly available.

In the case of SARS-CoV-2, past research shows evidence that normalizing wastewater concentrations of viral genetic material can improve correlations with reported COVID-19 cases across locations and somewhat over time \cite{duvallet2022}, making data processing an important component of WBE. Awareness of variation in PMMoV and research into how best to correct for its geographic trend may help to improve these correlations further, thereby enhancing the benefits and accuracy of wastewater-based epidemiology methods.

\clearpage
\begin{acknowledgement}

The authors thank Rasha Maal-Bared (CDM Smith), Lilly Pang and Bonita Lee (University of Alberta), Colleen Naughton (UC Merced), Claire Duvallet (Science for America, formerly Biobot Analytics), Scott Oleson (Center for Forecasting and Outbreak Analytics, CDC), Adam Smith (University of Southern California), Ben Yaffe (Verily Life Sciences, Inc.), Robert Delatolla (University of Ottawa), Dan Gerrity (University of Nevada), Katherine Crank (Southern Nevada Water Authority), and Allison Wheeler (Colorado Department of Public Health) for sharing their guidance and expertise on wastewater treatment and wastewater-based epidemiology. The authors also thank the wastewater treatment facilities collaborating with WastewaterSCAN for providing invaluable data for this research and the members of the Delphi research group (Carnegie Mellon University) for their ongoing support. 

Funding for ABB, ALB, and MKW was through a gift to Stanford University from the Sergey Brin Family Foundation.

This material is based upon work supported by the United States of America Department of Health and Human Services, Centers for Disease Control and Prevention, under award number NU38FT000005; and contract number 75D30123C1590. Any opinions, findings, and conclusions or recommendations expressed in this material are those of the author(s) and do not necessarily reflect the views of the United States of America Department of Health and Human Services, Centers for Disease Control and Prevention.

\end{acknowledgement}

\begin{suppinfo}
  
A complete list of R packages used in analysis, illustration of the component effects in the Bayesian median models, autocorrelation analyses, supplemental variance decomposition, tables for all model fits, site name abbreviations table, and site summary statistics are found in the supplementary information file. 

Code to reproduce analyses is organized into the \emph{pmmov} package in R. The code script to reproduce the figures in this manuscript (\emph{manu.R}), and the Stan model and fitting script for the Bayesian median models (\emph{pmmov.stan} and \emph{stanfit.R}) are also available free of charge. 
All files can be found at the accompanying GitHub repository (\href{https://github.com/aerosengart/pmmov-manu.git}{https://github.com/aerosengart/pmmov-manu.git}).

\end{suppinfo}

\clearpage
\bibliography{manu}

\end{document}

\newpage

\section{R Packages}
\label{sec:analysis}
The \emph{xlsx} package was used for additional data reading and writing \cite{dragulescu2020}. Data manipulation was done using the \emph{dplyr} and \emph{magrittr} packages \cite{wickham2023, wickham2022}. Zip codes for each site were found using the \emph{zipcodeR} package \cite{rozzi2021}. Fourier bases were constructed with the \emph{fda} package \cite{ramsay2023}. Linear regression models were fit with the \emph{stats} package \cite{r2021}. Quantile regression models were fit with the \emph{quantreg} package \cite{koenker2022} using the modified Barrodale and Roberts algorithm \cite{koenker1987, koenker1994}, and standard errors were estimated using the cluster-robust wild bootstrap with grouping defined by site \cite{hagemann2017}. The Bayesian models were fit with the \emph{rstan} package \cite{rstan2024}. Plots were made using the \emph{ggplot2}, \emph{ggrepel}, \emph{scales}, \emph{ggpubr}, and \emph{cowplot} packages \cite{wickham2016, slowikowski2021, wickham2023b, kassambara2020, wilke2020}. The \emph{sf} and \emph{concaveman} packages were also used for creating maps \cite{pebesma2018, pebesma2023, gombin2020}.

\section{Component Effects in Bayesian Median Models}
\label{subsec:components}
\begin{figure}[H]
  \centering
  \includegraphics[width=0.9\textwidth]{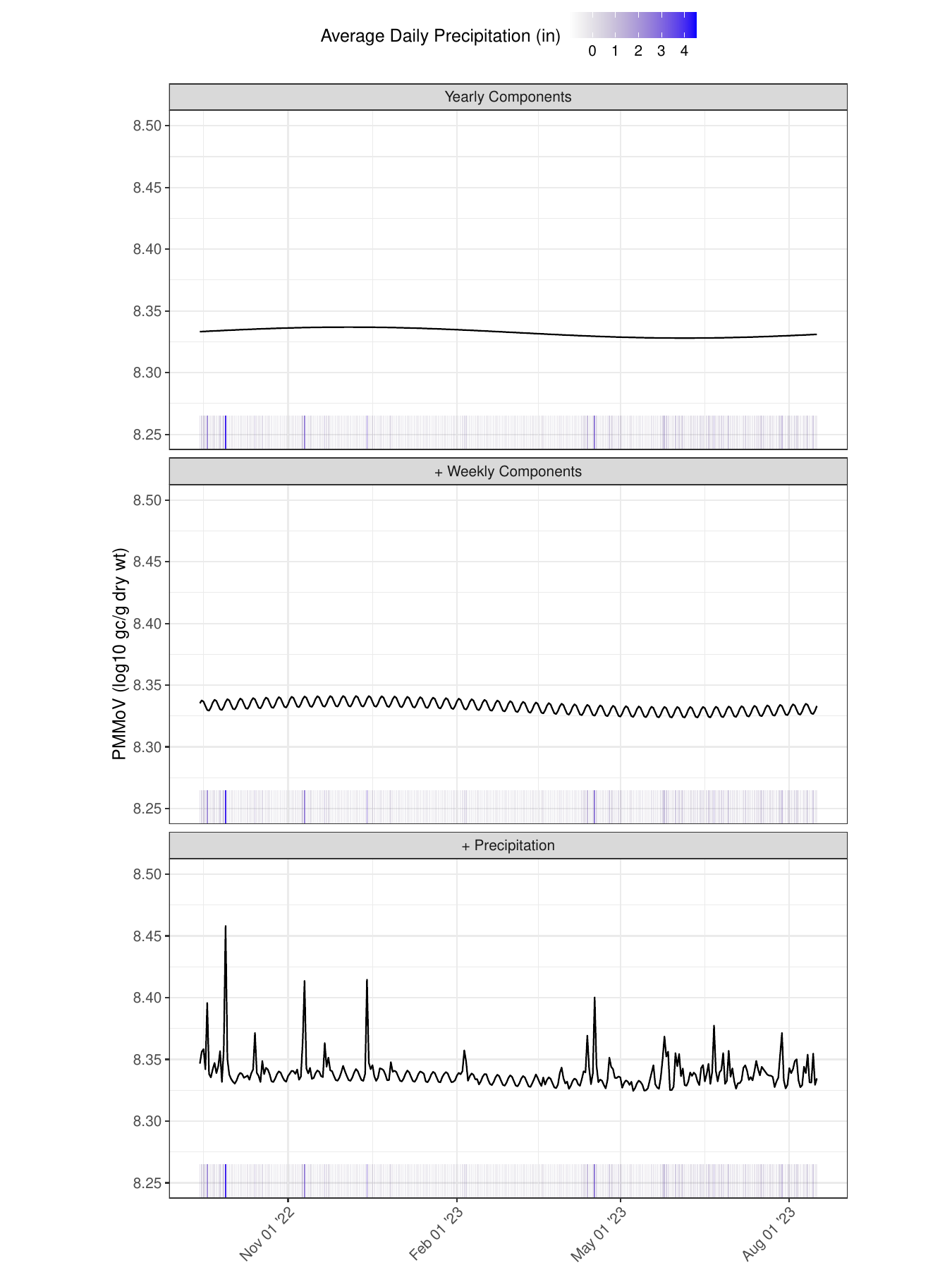}
  \caption{Predicted PMMoV concentration at Loxahatchee River Environmental Control District separated by component effects from one posterior sample. Rug shows average daily precipitation. (gc = gene copies; g = gram; wt = weight; in = inches)}
  \label{fig:components}
\end{figure}

To illustrate the effect of each component in the Bayesian median models, we take the Loxahatchee River Environmental Control District site in Florida as an example. We sample from the posterior distribution of the coefficients and use these values to construct the predicted PMMoV concentration by component. The top panel of Figure~\ref{fig:components} shows the effect of the yearly time components in addition to the intercept. The middle panel adds the effect of the weekly components, and the bottom panel adds precipitation. 

\section{Autocorrelation Analysis}
\begin{figure}[H]
  \centering
  \begin{subfigure}[t]{0.45\textwidth}
    \centering
    \includegraphics[width=\textwidth]{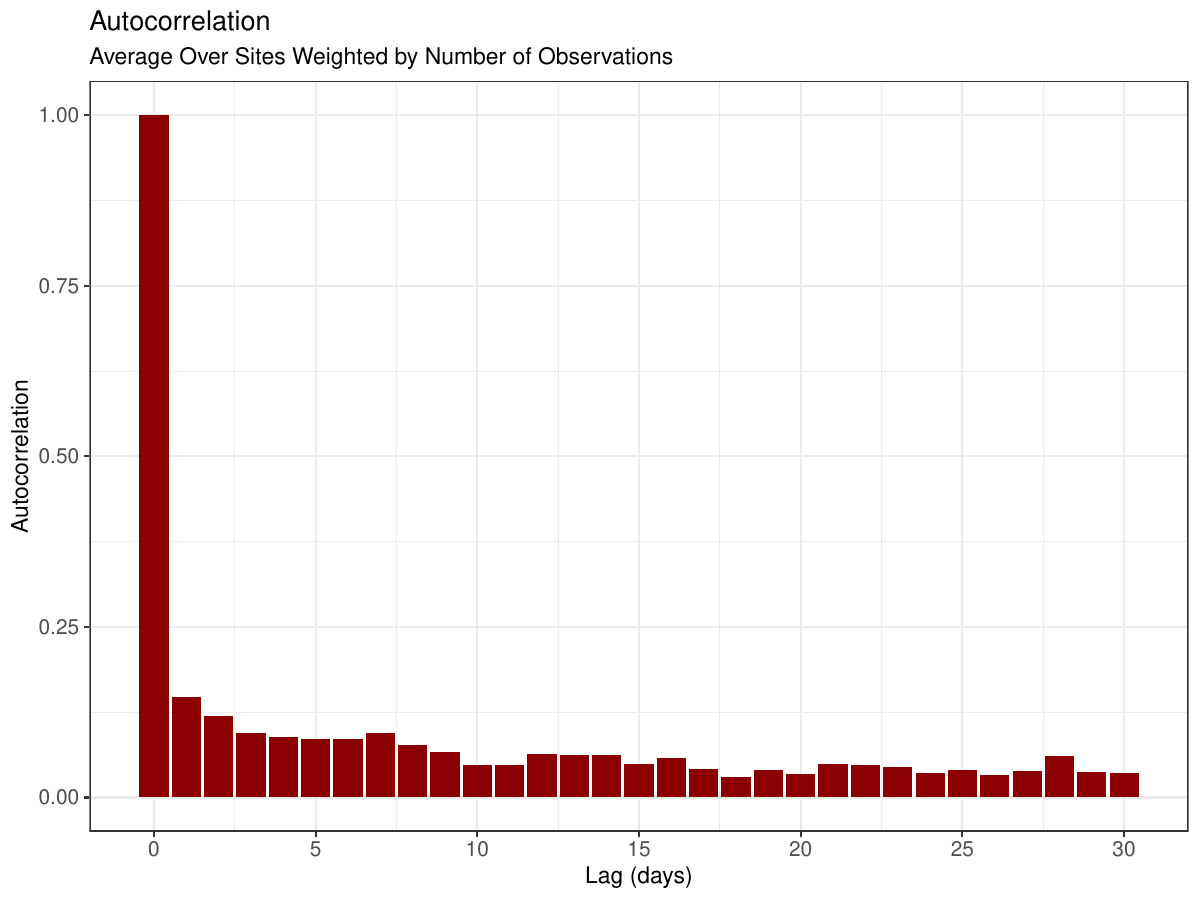}
    \caption{Autocorrelation of $\log_{10}$ PMMoV concentration across all sites exhibits a weekly pattern with higher autocorrelation values for weekly lags (7 days, 14 days, etc.).}
    \label{fig:autocorr-plot-data}
  \end{subfigure}
  \hfill
  \begin{subfigure}[t]{0.45\textwidth}
      \centering
      \includegraphics[width=\textwidth]{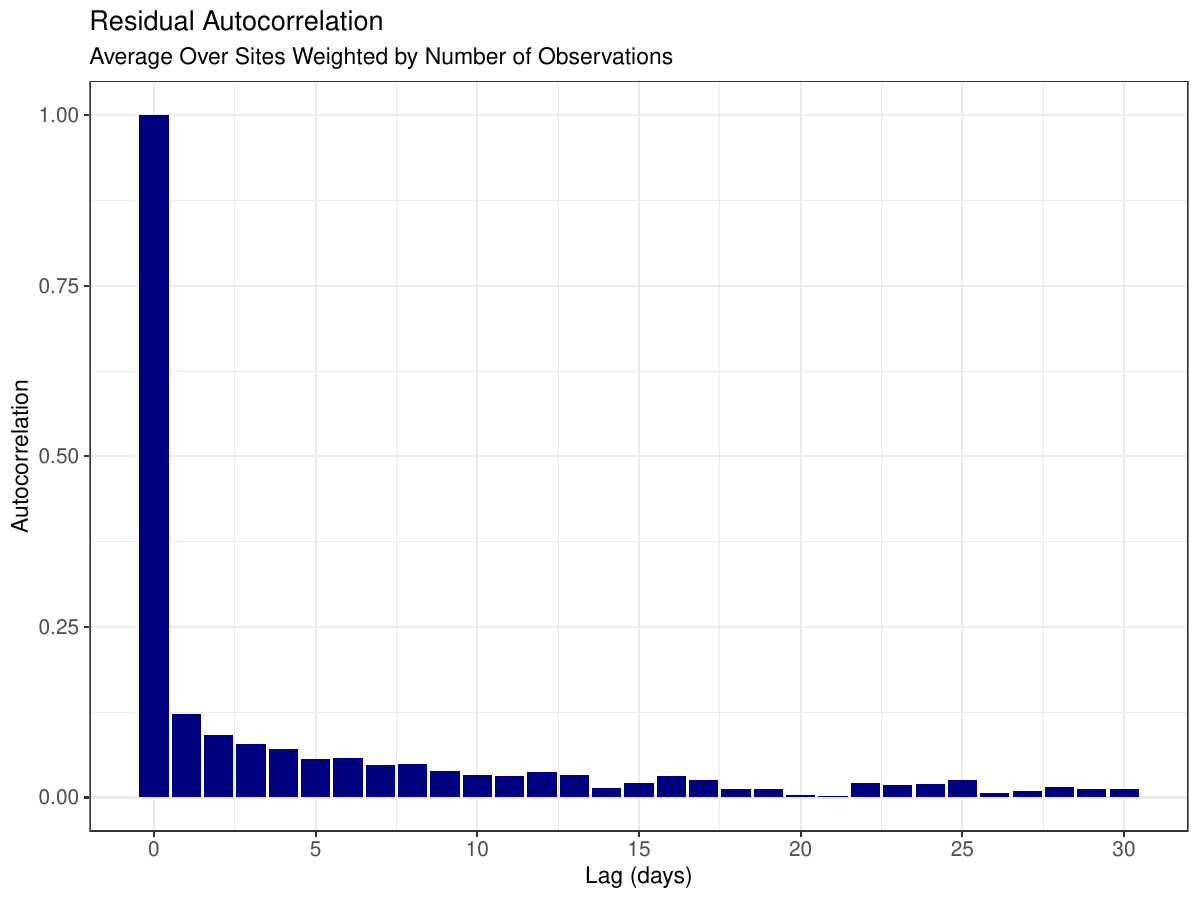}
      \caption{Autocorrelation of residuals from Bayesian median models for all sites.}
      \label{fig:autocorr-plot-resid}
  \end{subfigure}
  \caption{Summary autocorrelation across all sites.}
\end{figure}

Figure~\ref{fig:autocorr-plot-data} provides a general summary of the level of autocorrelation in the data. The weighted average autocorrelation is calculated by first augmenting the data by adding rows such that each site's time series is daily. Days with missing observations are filled with an \emph{NA} value. The within-site autocorrelation is then calculated in a similar way to the standard sample autocorrelation \cite{brockwell2006}; however, only pairs of observations for which there are data at time $t$ and time $t + k$ (for lag of $k$ days) are included in the calculations. If there are too few observations such that the lag $k$ autocorrelation cannot be calculated for a given site, then a value of $0$ is substituted. For each value of $k = 0, \dots, 30$, the weighted average of the lag $k$ autocorrelations for all sites is taken with the weights determined by the number of pairs of observations used to calculate each within-site autocorrelation. Figure~\ref{fig:autocorr-plot-resid} performs the same calculations with the residuals of the Bayesian median models, which include weekly components. There is a reduction in autocorrelation, especially at weekly lags (7 days, 14 days, etc.), when comparing the raw data to the models' residuals.

\section{Additional Variance Decomposition}
\label{subsec:add-var-decomp}
\begin{figure}[H]
  \centering
  \includegraphics[width=\textwidth]{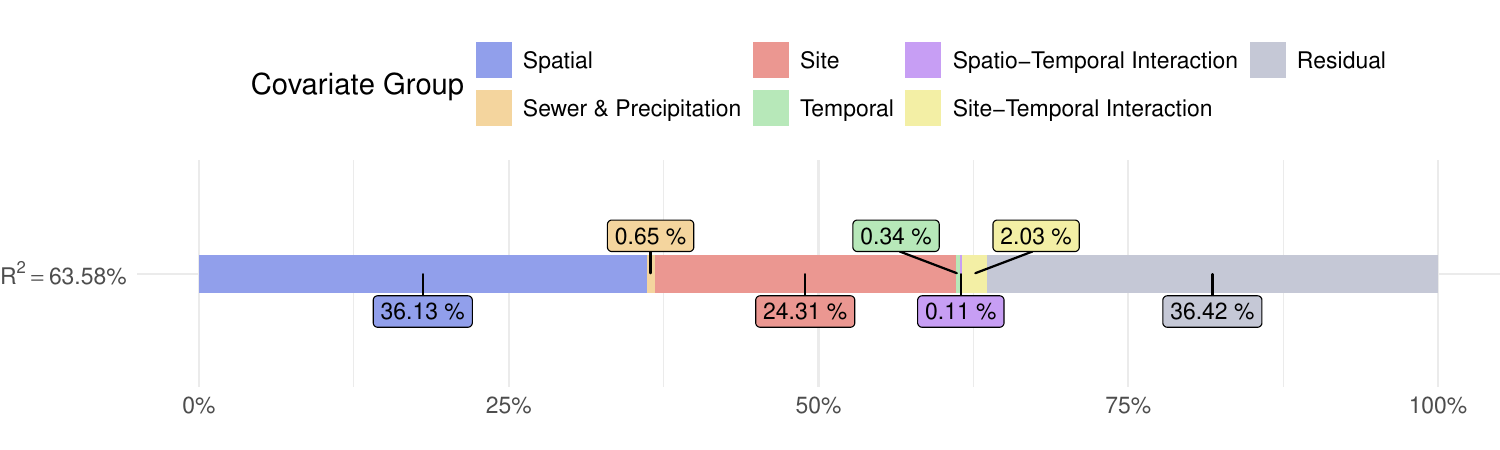}
  \caption{Location and site membership account for the majority of the variation in PMMoV concentration. The spatial components (latitude and longitude) account for the greatest portion at over 36\%. The temporal components, sewer system type, and precipitation altogether account for around 1\% of the variation. Site membership accounts for over 24\% of the variation that remains unexplained by the spatial, sewer, and precipitation variables. Site- and location-specific temporal features explain only 2.14\% of the remaining variation.}
  \label{fig:var-decomp-interactions}
\end{figure}

We performed an additional variance decomposition in the same way as in Section~\ref*{subsubsection:model-3} of the main text but included two additional groups of covariates that allowed for site- and location-specific temporal components through interaction terms:
\begin{equation}
  \label{eq:model-3b}
  \begin{split}
    \mathbb{E}\left[ \text{log}_{10} \text{PMMoV}_{i, t} \right] &= \beta_{0} + \beta_1 \cdot \text{lat}_i + \beta_2 \cdot \text{lng}_i + \beta_3 \cdot \text{prcp}_{i,t} + \beta_4 \cdot \text{sewer}_i \\
    &+ \beta_5 \cdot \text{prcp}_{i, t} \cdot \text{sewer}_i + \beta_6 \cdot \text{site ID}_i \\
    &+ \beta_7 \cdot \psi^{\text{sin}}_{7}(t) + \beta_8 \cdot \psi^{\text{cos}}_{7}(t) + \beta_{9} \cdot \psi^{\text{sin}}_{365.25}(t) + \beta_{10} \cdot \psi^{\text{cos}}_{365.25}(t) + \\
    &+ \beta_{11} \cdot \psi^{\text{sin}}_{7}(t) \cdot \text{lat}_i + \beta_{12} \cdot \psi^{\text{cos}}_{7}(t) \cdot \text{lat}_i \\
    &+ \beta_{13} \cdot \psi^{\text{sin}}_{365.25}(t) \cdot \text{lat}_i + \beta_{14} \cdot \psi^{\text{cos}}_{365.25}(t) \cdot \text{lat}_i \\
    &+ \beta_{15} \cdot \psi^{\text{sin}}_{7}(t) \cdot \text{lng}_i + \beta_{16} \cdot \psi^{\text{cos}}_{7}(t) \cdot \text{lng}_i \\
    &+ \beta_{17} \cdot \psi^{\text{sin}}_{365.25}(t) \cdot \text{lng}_i + \beta_{18} \cdot \psi^{\text{cos}}_{365.25}(t) \cdot \text{lng}_i \\
    &+ \beta_{19} \cdot \psi^{\text{sin}}_{7}(t) \cdot \text{site ID}_i + \beta_{20} \cdot \psi^{\text{cos}}_{7}(t) \cdot \text{site ID}_i \\
    &+ \beta_{21} \cdot \psi^{\text{sin}}_{365.25}(t) \cdot \text{site ID}_i + \beta_{22} \cdot \psi^{\text{cos}}_{365.25}(t) \cdot \text{site ID}_i
  \end{split}
\end{equation}
The order of covariate addition was: (i) latitude and longitude; (ii) precipitation, sewer system type, and their interaction; (iii) site indicator; (iv) the weekly and yearly time components; (v) interaction terms between latitude, longitude, and the temporal components; (vi) interaction terms between the site indicator and the temporal components.

\section{Model Fits}
\label{sec:model-fits}

\subsection{Simple Median Model}
\begin{table}[H]
    \centering
    \caption{Coefficient estimates, cluster-robust wild bootstrap standard errors, and associated $p$-values for the simple median model. Coefficients with bold $p$-values indicate statistical signficance at level $\alpha = 0.05$.}
    \label{tab:model-1}
      \begin{tabular}{l|lll}
                & Estimate & Std. Error & $p$-value \\ \hline
      Intercept & \hspace{2mm} $7.411$ & $0.236$ & $\mathbf{< 1.000 \times 10^{-15}}$ \\
      Lat. & $-9.764 \times 10^{-5}$ & $5.139 \times 10^{-3}$ & \hspace{3mm} $0.985$ \\
      Lng. & $-1.286 \times 10^{-2}$ & $1.067 \times 10^{-3}$ & $\mathbf{< 1.000 \times 10^{-15}}$
      \end{tabular}
  \end{table}

\subsection{Detailed Median Model}
\begin{table}[H]
  \centering
  \caption{Coefficient estimates, cluster-robust wild bootstrap standard errors, and associated $p$-values for the detailed median model. Bold $p$-values indicate statistical signficance at level $\alpha = 0.05$.}
  \label{tab:model-2}
  \begin{tabular}{l|lll}
                & Estimate                  & Std. Error            & $p$-value   \\ \hline
  Intercept & \hspace{2.3mm} $7.408$ & $0.251$ & $\mathbf{< 1.000 \times 10^{-15}}$ \\
  Lat. & $-2.462 \times 10^{-4}$ & $5.286 \times 10^{-3}$ & \hspace{3mm} $0.963$ \\
  Lng. & $-1.316 \times 10^{-2}$ & $1.178 \times 10^{-3}$ & $\mathbf{< 1.000 \times 10^{-15}}$ \\
  Avg. Prcp. & $-7.326 \times 10^{-2}$ & $9.875 \times 10^{-3}$ &
   \hspace{3mm} $\mathbf{1.217 \times 10^{-13}}$ \\
  Sewer & $-0.116$ & $7.454 \times 10^{-2}$ & \hspace{3mm} $0.119$ \\
  Avg. Prcp./Sewer & $-2.784 \times 10^{-2}$ & $5.801 \times 10^{-2}$ &    \hspace{3mm} $0.631$ \\
  $\psi_{7}^{\text{sin}}$ & \hspace{2.3mm} $2.375 \times 10^{-2}$ & $1.460 \times 10^{-2}$ & \hspace{3mm} $0.104$ \\
  $\psi_{7}^{\text{sin}}$ & $-1.613 \times 10^{-2}$ & $1.241 \times 10^{-2}$ & \hspace{3mm} $0.194$ \\
  $\psi_{365.25}^{\text{sin}}$ & $-0.157$ & $9.477 \times 10^{-2}$ & \hspace{3mm} $9.753 \times 10^{-2}$ \\
  $\psi_{365.25}^{\text{cos}}$ & $-0.343$ & $0.111$ & \hspace{3mm} $\mathbf{2.053 \times 10^{-3}}$ 
  \end{tabular}
\end{table}

\subsection{Variance Decomposition Model}
\begin{figure}[H]
  \centering
  \includegraphics[width=\textwidth]{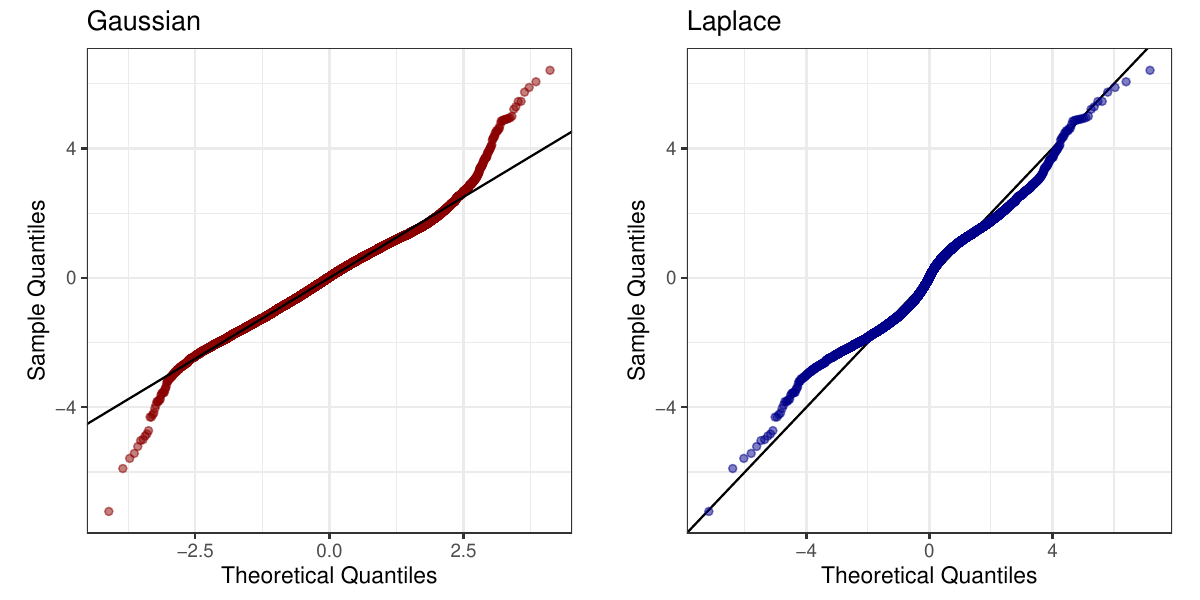}
  \caption{Sample quantiles of standardized residuals of the variance decomposition model against theoretical quantiles of standard Gaussian distribution (left) and standard Laplace distribution (right).}
  \label{fig:qq-plot}
\end{figure}

Coefficient estimates for the variance decomposition model are omitted due to the departure from the assumed Gaussianity of errors (Figure~\ref{fig:qq-plot}). Analysis of $R^2$ remains reasonable as it does not rely upon distributional assumptions.

\subsubsection{Bayesian Median Models}
\begin{tiny}
\begin{longtable}{|l|llll|ll|}
      \caption{Mean, Monte Carlo standard error, median, 95\% credible interval, effective sample size, and $\hat{R}$ for the scale parameter $\sigma$ calculated over 4,000 posterior samples.}
      \label{tab:sigma} \\ \hline
      Site                        & Mean    & MCSE                   & Median  & 95\% CI                   & ESS                   & R Hat   \\ \hline
      \endhead
      \hline
      \endfoot 
      Akron                       & $0.166$                & $4.176 \times 10^{-4}$ & $0.164$                & $(0.133, 0.208)$                               & $2.138 \times 10^{3}$ & $1.000$ \\
Altamonte Springs           & $0.145$                & $3.105 \times 10^{-4}$ & $0.144$                & $(0.120, 0.176)$                               & $2.009 \times 10^{3}$ & $1.002$ \\
Ann Arbor                   & $0.207$                & $3.449 \times 10^{-4}$ & $0.206$                & $(0.178, 0.239)$                               & $2.032 \times 10^{3}$ & $1.002$ \\
Aquia                       & $8.175 \times 10^{-2}$ & $2.089 \times 10^{-4}$ & $8.111 \times 10^{-2}$ & $(6.436 \times 10^{-2}, 0.103)$                & $2.196 \times 10^{3}$ & $1.002$ \\
Archie Elledge              & $0.223$                & $4.127 \times 10^{-4}$ & $0.222$                & $(0.188, 0.261)$                               & $2.066 \times 10^{3}$ & $1.002$ \\
Bangor                      & $0.334$                & $1.321 \times 10^{-3}$ & $0.325$                & $(0.226, 0.489)$                               & $2.681 \times 10^{3}$ & $1.001$ \\
Bayshore                    & $0.163$                & $6.060 \times 10^{-4}$ & $0.162$                & $(0.121, 0.218)$                               & $1.651 \times 10^{3}$ & $1.001$ \\
Big Creek                   & $0.215$                & $3.410 \times 10^{-4}$ & $0.214$                & $(0.185, 0.248)$                               & $2.230 \times 10^{3}$ & $1.001$ \\
\end{longtable}
\end{tiny}
\begin{tiny}
    \begin{longtable}{|l|llll|ll|}
      \caption{Mean, Monte Carlo standard error, median, 95\% credible interval, effective sample size, and $\hat{R}$ for intercept calculated over 4,000 posterior samples. Bolded intervals do not contain 0.}
      \label{tab:alpha} \\ \hline
      Site                        & Mean    & MCSE                   & Median  & 95\% CI                   & ESS                   & R Hat   \\ \hline
      \endhead
      \hline
      \endfoot 
      Akron                       & $8.487$ & $4.242 \times 10^{-4}$ & $8.487$ & $\mathbf{(8.452, 8.523)}$ & $1.874 \times 10^{3}$ & $1.000$ \\
Altamonte Springs           & $8.489$ & $4.079 \times 10^{-4}$ & $8.489$ & $\mathbf{(8.454, 8.522)}$ & $1.790 \times 10^{3}$ & $1.003$ \\
Ann Arbor                   & $8.654$ & $3.908 \times 10^{-4}$ & $8.655$ & $\mathbf{(8.618, 8.688)}$ & $2.128 \times 10^{3}$ & $1.001$ \\
Aquia                       & $8.635$ & $2.441 \times 10^{-4}$ & $8.635$ & $\mathbf{(8.613, 8.659)}$ & $2.191 \times 10^{3}$ & $1.003$ \\
Archie Elledge              & $8.602$ & $4.779 \times 10^{-4}$ & $8.602$ & $\mathbf{(8.560, 8.645)}$ & $2.129 \times 10^{3}$ & $1.002$ \\
Bangor                      & $7.892$ & $1.964 \times 10^{-3}$ & $7.884$ & $\mathbf{(7.725, 8.083)}$ & $2.208 \times 10^{3}$ & $1.001$ \\
Bayshore                    & $8.493$ & $5.867 \times 10^{-4}$ & $8.494$ & $\mathbf{(8.437, 8.545)}$ & $2.177 \times 10^{3}$ & $1.000$ \\
Big Creek                   & $8.641$ & $3.836 \times 10^{-4}$ & $8.641$ & $\mathbf{(8.608, 8.674)}$ & $1.935 \times 10^{3}$ & $1.000$ \\
\end{longtable}
    \end{tiny}
\begin{tiny}
    \begin{longtable}{|l|llll|ll|}
      \caption{Mean, Monte Carlo standard error, median, 95\% credible interval, effective sample size, and $\hat{R}$ for coefficient on precipitation calculated over 4,000 posterior samples. Bolded intervals do not contain 0.}
      \label{tab:prcp} \\ \hline
      Site                        & Mean    & MCSE                   & Median  & 95\% CI                   & ESS                   & R Hat   \\ \hline
      \endhead
      \hline
      \endfoot 
      Akron                       & $-1.809 \times 10^{-2}$ & $3.310 \times 10^{-4}$ & $-1.606 \times 10^{-2}$ & $(-6.035 \times 10^{-2}, 1.032 \times 10^{-2})$           & $2.913 \times 10^{3}$ & $1.000$ \\
Altamonte Springs           & $1.808 \times 10^{-2}$  & $3.978 \times 10^{-4}$ & $1.840 \times 10^{-2}$  & $(-2.094 \times 10^{-2}, 5.497 \times 10^{-2})$           & $2.270 \times 10^{3}$ & $1.000$ \\
Ann Arbor                   & $1.240 \times 10^{-2}$  & $3.105 \times 10^{-4}$ & $1.183 \times 10^{-2}$  & $(-1.707 \times 10^{-2}, 4.394 \times 10^{-2})$           & $2.459 \times 10^{3}$ & $1.000$ \\
Aquia                       & $-6.988 \times 10^{-3}$ & $2.434 \times 10^{-4}$ & $-6.167 \times 10^{-3}$ & $(-3.049 \times 10^{-2}, 1.498 \times 10^{-2})$           & $2.080 \times 10^{3}$ & $1.000$ \\
Archie Elledge              & $-2.059 \times 10^{-2}$ & $3.537 \times 10^{-4}$ & $-2.030 \times 10^{-2}$ & $(-5.469 \times 10^{-2}, 1.106 \times 10^{-2})$           & $2.312 \times 10^{3}$ & $1.000$ \\
Bangor                      & $7.064 \times 10^{-3}$  & $1.313 \times 10^{-3}$ & $6.212 \times 10^{-3}$  & $(-0.144, 0.145)$                                         & $2.834 \times 10^{3}$ & $1.002$ \\
Bayshore                    & $-7.131 \times 10^{-3}$ & $6.881 \times 10^{-4}$ & $-4.842 \times 10^{-3}$ & $(-8.109 \times 10^{-2}, 6.818 \times 10^{-2})$           & $2.913 \times 10^{3}$ & $1.001$ \\
Big Creek                   & $3.329 \times 10^{-4}$  & $2.812 \times 10^{-4}$ & $8.591 \times 10^{-4}$  & $(-3.681 \times 10^{-2}, 3.095 \times 10^{-2})$           & $3.227 \times 10^{3}$ & $1.000$ \\
\end{longtable}
\end{tiny}
\begin{tiny}
    \begin{longtable}{|l|llll|ll|}
      \caption{Mean, Monte Carlo standard error, median, 95\% credible interval, effective sample size, and $\hat{R}$ for coefficient on $\psi_7^{\text{sin}}$ (weekly basis function) calculated over 4,000 posterior samples. Bolded intervals do not contain 0.}
      \label{tab:sinwk} \\ \hline
      Site                        & Mean    & MCSE                   & Median  & 95\% CI                   & ESS                   & R Hat   \\ \hline
      \endhead
      \hline
      \endfoot 
      Akron                       & $-4.483 \times 10^{-3}$ & $3.441 \times 10^{-4}$ & $-3.444 \times 10^{-3}$ & $(-3.873 \times 10^{-2}, 2.945 \times 10^{-2})$           & $2.389 \times 10^{3}$ & $1.001$ \\
Altamonte Springs           & $-1.187 \times 10^{-2}$ & $3.119 \times 10^{-4}$ & $-1.105 \times 10^{-2}$ & $(-4.302 \times 10^{-2}, 1.580 \times 10^{-2})$           & $2.305 \times 10^{3}$ & $1.001$ \\
Ann Arbor                   & $2.803 \times 10^{-3}$  & $3.088 \times 10^{-4}$ & $1.949 \times 10^{-3}$  & $(-2.912 \times 10^{-2}, 3.601 \times 10^{-2})$           & $2.666 \times 10^{3}$ & $1.000$ \\
Aquia                       & $-7.338 \times 10^{-4}$ & $1.924 \times 10^{-4}$ & $-4.390 \times 10^{-4}$ & $(-2.008 \times 10^{-2}, 1.856 \times 10^{-2})$           & $2.637 \times 10^{3}$ & $1.003$ \\
Archie Elledge              & $1.102 \times 10^{-2}$  & $4.660 \times 10^{-4}$ & $9.617 \times 10^{-3}$  & $(-3.427 \times 10^{-2}, 5.661 \times 10^{-2})$           & $2.350 \times 10^{3}$ & $1.001$ \\
Bangor                      & $-2.961 \times 10^{-2}$ & $1.507 \times 10^{-3}$ & $-2.264 \times 10^{-2}$ & $(-0.194, 0.120)$                                         & $2.564 \times 10^{3}$ & $1.000$ \\
Bayshore                    & $-8.444 \times 10^{-4}$ & $3.996 \times 10^{-4}$ & $-4.477 \times 10^{-4}$ & $(-4.642 \times 10^{-2}, 4.314 \times 10^{-2})$           & $3.077 \times 10^{3}$ & $0.999$ \\
Big Creek                   & $3.058 \times 10^{-2}$  & $5.391 \times 10^{-4}$ & $2.961 \times 10^{-2}$  & $(-1.619 \times 10^{-2}, 8.457 \times 10^{-2})$           & $2.252 \times 10^{3}$ & $1.001$ \\
\end{longtable}
\end{tiny}
\begin{tiny}
    \begin{longtable}{|l|llll|ll|}
      \caption{Mean, Monte Carlo standard error, median, 95\% credible interval, effective sample size, and $\hat{R}$ for coefficient on $\psi_7^{\text{cos}}$ (weekly basis function) calculated over 4,000 posterior samples. Bolded intervals do not contain 0.}
      \label{tab:coswk} \\ \hline
      Site                        & Mean    & MCSE                   & Median  & 95\% CI                   & ESS                   & R Hat   \\ \hline
      \endhead
      \hline
      \endfoot 
      Akron                       & $6.893 \times 10^{-3}$  & $3.473 \times 10^{-4}$ & $5.073 \times 10^{-3}$  & $(-2.468 \times 10^{-2}, 4.313 \times 10^{-2})$           & $2.460 \times 10^{3}$ & $1.000$ \\
Altamonte Springs           & $-1.886 \times 10^{-2}$ & $3.562 \times 10^{-4}$ & $-1.796 \times 10^{-2}$ & $(-5.711 \times 10^{-2}, 1.246 \times 10^{-2})$           & $2.609 \times 10^{3}$ & $1.002$ \\
Ann Arbor                   & $-8.505 \times 10^{-3}$ & $4.353 \times 10^{-4}$ & $-6.760 \times 10^{-3}$ & $(-5.155 \times 10^{-2}, 2.963 \times 10^{-2})$           & $2.076 \times 10^{3}$ & $1.000$ \\
Aquia                       & $2.632 \times 10^{-2}$  & $2.109 \times 10^{-4}$ & $2.578 \times 10^{-2}$  & $\mathbf{(3.697 \times 10^{-3}, 5.131 \times 10^{-2})}$   & $3.340 \times 10^{3}$ & $1.000$ \\
Archie Elledge              & $-3.569 \times 10^{-2}$ & $4.994 \times 10^{-4}$ & $-3.543 \times 10^{-2}$ & $(-8.163 \times 10^{-2}, 5.958 \times 10^{-3})$           & $2.124 \times 10^{3}$ & $1.000$ \\
Bangor                      & $6.378 \times 10^{-2}$  & $1.534 \times 10^{-3}$ & $6.046 \times 10^{-2}$  & $(-9.286 \times 10^{-2}, 0.228)$                          & $2.681 \times 10^{3}$ & $1.000$ \\
Bayshore                    & $-6.056 \times 10^{-3}$ & $5.074 \times 10^{-4}$ & $-2.655 \times 10^{-3}$ & $(-6.703 \times 10^{-2}, 4.452 \times 10^{-2})$           & $2.955 \times 10^{3}$ & $1.001$ \\
Big Creek                   & $1.450 \times 10^{-2}$  & $4.693 \times 10^{-4}$ & $1.121 \times 10^{-2}$  & $(-2.453 \times 10^{-2}, 6.552 \times 10^{-2})$           & $2.432 \times 10^{3}$ & $1.001$ \\
\end{longtable}
\end{tiny}
\begin{tiny}
    \begin{longtable}{|l|llll|ll|}
      \caption{Mean, Monte Carlo standard error, median, 95\% credible interval, effective sample size, and $\hat{R}$ for coefficient on $\psi_{365.25}^{\text{sin}}$ (yearly basis function) calculated over 4,000 posterior samples. Bolded intervals do not contain 0.}
      \label{tab:sinyr} \\ \hline
      Site                        & Mean    & MCSE                   & Median  & 95\% CI                   & ESS                   & R Hat   \\ \hline
      \endhead
      \hline
      \endfoot 
      Akron                       & $-1.503 \times 10^{-2}$ & $4.690 \times 10^{-4}$ & $-1.377 \times 10^{-2}$ & $(-6.075 \times 10^{-2}, 2.748 \times 10^{-2})$           & $2.240 \times 10^{3}$ & $1.000$ \\
Altamonte Springs           & $-4.926 \times 10^{-2}$ & $2.951 \times 10^{-4}$ & $-4.964 \times 10^{-2}$ & $\mathbf{(-7.947 \times 10^{-2}, -1.724 \times 10^{-2})}$ & $2.867 \times 10^{3}$ & $1.001$ \\
Ann Arbor                   & $-1.395 \times 10^{-2}$ & $3.048 \times 10^{-4}$ & $-1.323 \times 10^{-2}$ & $(-4.809 \times 10^{-2}, 1.855 \times 10^{-2})$           & $3.013 \times 10^{3}$ & $0.999$ \\
Aquia                       & $-4.564 \times 10^{-2}$ & $1.887 \times 10^{-4}$ & $-4.610 \times 10^{-2}$ & $\mathbf{(-6.867 \times 10^{-2}, -2.092 \times 10^{-2})}$ & $4.134 \times 10^{3}$ & $1.000$ \\
Archie Elledge              & $-4.973 \times 10^{-2}$ & $4.214 \times 10^{-4}$ & $-4.885 \times 10^{-2}$ & $\mathbf{(-9.789 \times 10^{-2}, -4.737 \times 10^{-3})}$ & $3.139 \times 10^{3}$ & $1.001$ \\
Bangor                      & $0.115$                 & $3.116 \times 10^{-3}$ & $9.789 \times 10^{-2}$  & $(-0.143, 0.439)$                                         & $2.254 \times 10^{3}$ & $1.000$ \\
Bayshore                    & $-0.125$                & $1.056 \times 10^{-3}$ & $-0.126$                & $\mathbf{(-0.216, -2.300 \times 10^{-2})}$                & $2.111 \times 10^{3}$ & $1.002$ \\
Big Creek                   & $2.377 \times 10^{-2}$  & $3.263 \times 10^{-4}$ & $2.332 \times 10^{-2}$  & $(-8.908 \times 10^{-3}, 6.020 \times 10^{-2})$           & $3.002 \times 10^{3}$ & $0.999$ \\
\end{longtable}
\end{tiny}
\begin{tiny}
    \begin{longtable}{|l|llll|ll|}
      \caption{Mean, Monte Carlo standard error, median, 95\% credible interval, effective sample size, and $\hat{R}$ for coefficient on $\psi_{365.25}^{\text{cos}}$ (yearly basis function) calculated over 4,000 posterior samples. Bolded intervals do not contain 0.}
      \label{tab:cosyr} \\ \hline
      Site                        & Mean    & MCSE                   & Median  & 95\% CI                   & ESS                   & R Hat   \\ \hline
      \endhead
      \hline
      \endfoot 
      Akron                       & $3.438 \times 10^{-2}$  & $4.013 \times 10^{-4}$ & $3.432 \times 10^{-2}$  & $(-1.282 \times 10^{-3}, 7.545 \times 10^{-2})$           & $2.470 \times 10^{3}$ & $1.000$ \\
Altamonte Springs           & $-7.636 \times 10^{-3}$ & $2.961 \times 10^{-4}$ & $-6.691 \times 10^{-3}$ & $(-3.699 \times 10^{-2}, 1.877 \times 10^{-2})$           & $2.188 \times 10^{3}$ & $1.001$ \\
Ann Arbor                   & $-4.687 \times 10^{-2}$ & $4.478 \times 10^{-4}$ & $-4.625 \times 10^{-2}$ & $\mathbf{(-8.826 \times 10^{-2}, -7.572 \times 10^{-3})}$ & $2.085 \times 10^{3}$ & $1.001$ \\
Aquia                       & $1.300 \times 10^{-4}$  & $1.955 \times 10^{-4}$ & $1.362 \times 10^{-4}$  & $(-2.108 \times 10^{-2}, 2.161 \times 10^{-2})$           & $3.041 \times 10^{3}$ & $1.000$ \\
Archie Elledge              & $1.273 \times 10^{-3}$  & $4.101 \times 10^{-4}$ & $5.586 \times 10^{-4}$  & $(-4.204 \times 10^{-2}, 4.537 \times 10^{-2})$           & $2.686 \times 10^{3}$ & $0.999$ \\
Bangor                      & $0.389$                 & $4.429 \times 10^{-3}$ & $0.386$                 & $\mathbf{(1.794 \times 10^{-2}, 0.772)}$                  & $1.809 \times 10^{3}$ & $1.001$ \\
Bayshore                    & $-7.737 \times 10^{-2}$ & $9.617 \times 10^{-4}$ & $-7.778 \times 10^{-2}$ & $(-0.165, 3.147 \times 10^{-3})$                          & $1.991 \times 10^{3}$ & $1.001$ \\
Big Creek                   & $-1.479 \times 10^{-2}$ & $3.266 \times 10^{-4}$ & $-1.372 \times 10^{-2}$ & $(-4.808 \times 10^{-2}, 1.414 \times 10^{-2})$           & $2.346 \times 10^{3}$ & $1.000$ \\
\end{longtable}
\end{tiny}

\begin{tiny}
    \begin{longtable}{|l|l|}
      \caption{Site name abbreviations. Sites ordered alphabetically by abbreviation.}
      \label{tab:abbrev} \\ \hline
        Site Name                                                                            & Abbreviation               \\ \hline
        \endhead
        \hline
        \endfoot 
        {[}Fremont Basin{]} -  Raymond A. Boege Alvarado WWTP                                 & Boege Alvarado (Fremont)    \\
        {[}Newark Basin{]} - Raymond A. Boege Alvarado WWTP                                   & Boege Alvarado (Newark)     \\
        {[}Union City Basin{]} - Raymond A. Boege Alvarado WWTP                               & Boege Alvarado (Union City) \\
        Akron Water Reclamation Facility                                                      & Akron                       \\
        Altamonte Springs Regional Water Reclamation Facility                                 & Altamonte Springs           \\
        Aquia Wastewater Treatment Facility                                                   & Aquia                       \\
        Archie Elledge WWTP                                                                   & Archie Elledge              \\
        Bayshore Regional Sewerage Authority                                                  & Bayshore                    \\
\end{longtable}
    \end{tiny}

\begin{landscape}

\begin{tiny}
    \begin{longtable}{|l|l|r|r|cccc|cccc|}
    \caption{Site summary statistics. Sites ordered alphabetically by abbreviation. PMMoV values are in $\log_{10}$ gene copies per gram dry weight. Precipitation values are in inches. (pop = population; obs = observations; prcp = precipitation)}
    \label{tab:summary-sites} \\ \hline
    Site & Location  & Pop Served & Num of Obs & 
    \multicolumn{1}{|p{1cm}|}{\centering Min \\ PMMoV} & 
    \multicolumn{1}{|p{1cm}|}{\centering Max \\ PMMoV} & 
    \multicolumn{1}{|p{1cm}|}{\centering Med \\ PMMoV} & 
    \multicolumn{1}{|p{1cm}|}{\centering Mean \\ PMMoV}& 
    \multicolumn{1}{|p{1cm}|}{\centering Min \\ Prcp} & 
    \multicolumn{1}{|p{1cm}|}{\centering Max \\ Prcp}& 
    \multicolumn{1}{|p{1cm}|}{\centering Med \\ Prcp} & 
    \multicolumn{1}{|p{1cm}|}{\centering Mean \\ Prcp}\\ \hline
    \endhead
    \hline
    \endfoot 
Akron                       & Akron, OH                    & 365000      & 85           & 7.37       & 9.31       & 8.49       & 8.49       & 0.00       & 1.26       & 0.01       & 0.10       \\
Altamonte Springs           & Altamonte Springs, FL        & 95000       & 117          & 8.03       & 9.16       & 8.49       & 8.49       & 0.00       & 1.72       & 0.00       & 0.12       \\
Ann Arbor                   & Ann Arbor, MI                & 125000      & 174          & 6.63       & 9.70       & 8.67       & 8.60       & 0.00       & 0.94       & 0.01       & 0.08       \\
Aquia                       & Stafford, VA                 & 100000      & 73           & 7.96       & 8.87       & 8.64       & 8.63       & 0.00       & 0.56       & 0.01       & 0.06       \\
Archie Elledge              & Winston-Salem, NC            & 92000       & 139          & 7.80       & 9.77       & 8.61       & 8.62       & 0.00       & 1.80       & 0.00       & 0.15       \\
Bangor                      & Bangor, ME                   & 40000       & 30           & 6.45       & 8.55       & 8.09       & 7.91       & 0.00       & 1.94       & 0.02       & 0.19       \\
Bayshore                    & Union Beach, NJ              & 100000      & 48           & 7.97       & 9.22       & 8.47       & 8.48       & 0.00       & 2.22       & 0.00       & 0.17       \\
Big Creek                   & Roswell, GA                  & 189593      & 173          & 7.69       & 9.84       & 8.63       & 8.63       & 0.00       & 3.25       & 0.01       & 0.15       \\
\end{longtable}
\end{tiny}

\end{landscape}

\clearpage
\bibliography{manu}